\documentclass{article}

\usepackage[margin=1.5in]{geometry}

\usepackage[utf8]{inputenc} 
\usepackage[T1]{fontenc}    
\usepackage{hyperref}       
\usepackage{url}            
\usepackage{booktabs}       
\usepackage{amsfonts}       
\usepackage{nicefrac}       
\usepackage{microtype}      
\usepackage{lipsum}		
\usepackage{graphicx}
\usepackage{natbib}
\usepackage{doi}

\usepackage{authblk}

\usepackage{listings}
\usepackage{xcolor}
\usepackage{xspace}
\usepackage{subcaption}
\usepackage{amsthm}
\usepackage{textgreek}
\usepackage{multirow}
\usepackage[linesnumbered,ruled,vlined]{algorithm2e}
\usepackage{paralist}
\usepackage{enumitem}
\usepackage{amsmath}
\usepackage{amsfonts} 
\usepackage{tabularx}

\SetKwInput{KwInput}{Input}                
\SetKwInput{KwOutput}{Output}             
\definecolor{steelblue}{rgb}{0.27, 0.51, 0.71} 

\SetCommentSty{mycommfont}

\title{Automatic Reuse, Adaption, and Execution of Simulation Experiments via Provenance Patterns}
\date{} 	

\author[1]{Pia Wilsdorf\thanks{pia.wilsdorf@uni-rostock.de}}
\author[1]{Anja Wolpers\thanks{anja.wolpers@uni-rostock.de}}
\author[2]{Jason Hilton\thanks{J.D.Hilton@soton.ac.uk}}
\author[1]{Fiete Haack\thanks{fiete.haack@uni-rostock.de}}
\author[1]{Adelinde M. Uhrmacher\thanks{adelinde.uhrmacher@uni-rostock.de}}
\affil[1]{Institute for Visual and Analytic Computing, University of Rostock, Germany}
\affil[2]{Department of Social Statistics and Demography, University of Southampton, UK}

\begin{document}

\maketitle

\begin{abstract}
	Simulation experiments are typically conducted repeatedly during the model development process, for example, to re-validate if a behavioral property still holds after several model changes.
	Approaches for automatically reusing and generating simulation experiments can support modelers in conducting simulation studies in a more systematic and effective manner.
	They rely on explicit experiment specifications and, so far, on user interaction for initiating the reuse.
	Thereby, they are constrained to support the reuse of simulation experiments in a specific setting.
	Our approach now goes one step further by automatically identifying and adapting the experiments to be reused for a variety of scenarios.
	To achieve this, we exploit provenance graphs of simulation studies, which provide valuable information about the previous modeling and experimenting activities, and contain meta-information about the different entities that were used or produced during the simulation study.
	We define provenance patterns and associate them with a semantics, which allows us to interpret the different activities, and construct transformation rules for provenance graphs.
	Our approach is implemented in a Reuse and Adapt framework for Simulation Experiments (RASE) which can interface with various modeling and simulation tools.
	In the case studies, we demonstrate the utility of our framework for a) the repeated sensitivity analysis of an agent-based model of migration routes, and b) the cross-validation of two models of a cell signaling pathway.
\end{abstract}

\section{Introduction}
Simulation experiments reveal important information about the behavior of a model.
Therefore, a wide variety of simulation experiments are conducted during a simulation study~\cite{Balci2012lifecycle, Sargent2013verification}. 
Automatically generating and executing simulation experiments allows simulation studies to be conducted in an easier and more systematic manner.
One option presents the goal-directed reuse of simulation experiments.     
In~\cite{Peng2017reusing, Peng2016reusing}, statistical model checking experiments~\cite{Agha2918survey} were reused to check whether the composition or extension of simulation models still exhibits certain behavioral properties. 
The tested properties can be interpreted as requirements, specifying the expected behavior of the simulation model~\cite{Ruscheinski2019artifact}, or as hypotheses to be tested in the development of a simulation model~\cite{Lorig2019hypothesis}. 
In the area of cardiac cellular electrophysiology, simulation experiments have been automatically reused to compare different model variants specified in CellML to assess their underlying hypotheses and their validity~\cite{Cooper2016cardiac}. 
Other approaches focus on the reuse of a simulation experiment's results (outputs) for settings in which experiments are performed repeatedly with the same simulation model, e.g., with various parameter configurations, and aim to increase computational efficiency by avoiding the execution of simulation experiments~\cite{Feng2017green}. 

In the above approaches, the type of simulation experiment, and/or kind of simulation model (including the modeling formalisms) have been constrained to support an automatic reuse of simulation experiments in a specific setting. 
However, in general various simulation experiments tend to be conducted repeatably with different variants of the simulation model during its development, and thus the repetition of simulation experiments forms a salient feature of the modeling and simulation life cycle.  

To approach the question of how to support an automatic reuse of simulation experiments more generally while conducting simulation studies, necessary ingredients and accessible information sources need to be identified.  
A prerequisite for the reuse of simulation experiments is a clear separation of concerns between model, simulator, and simulation experiment~\cite{Zeigler1984multifacetted}. 
In addition, simulation experiments need to be explicitly specified to be accessible and reusable. 
Over the last two decades, various approaches have been developed that allow an explicit specification of simulation experiments. 
To those belong, model-based approaches such as~\cite{Teran2015model}, 
domain-specific languages such as SESSL~\cite{Ewald2014sessl}, or standardized formats such as SED-ML~\cite{Waltemath2011reproducible}.  
Only if simulation experiments are explicitly specified, they can be automatically interpreted. 
Their interpretation is facilitated by schemas~\cite{Wilsdorf2019simulation} for the different types of experiments, possibly complemented by ontologies about their various roles~\cite{Ruscheinski2019artifact} and designs~\cite{Sanchez2020work}.

Also the past contains valuable information that can be exploited in a variation of Santayana's phrase ``Those who cannot remember the past are condemned to repeat it.'': Those who can remember and interpret the past can effectively plan the steps ahead which might include, in the case of simulation studies, a deliberate repetition of steps.  
Provenance information about simulation models reveals crucial information about a simulation model's past in terms of how a model has been developed. 
This includes information sources as well as activities, such as the conduction of simulation experiments that contributed to its development~\cite{Ruscheinski2017provenance}. 
Provenance information may be used to relate information sources, activities and generated entities, within and beyond individual simulation models thus forming entire families of models~\cite{Budde2021relating}. 
Here, we will pursue the question of how to automatically detect new experiments to be reused based on what has been done before~\cite{Wilsdorf2020conducting}. 
The reuse of simulation experiments refers to the reuse of a simulation experiment specification, which is then adapted and executed. 

As the central building block of our approach, we define typical patterns that can be observed in the provenance graph of simulation studies, and associate them with a semantics.
Based on the patterns we specify rules that automatically identify experiments to reuse, and then adapt, generate, and execute a new experiment. 
Updates of the provenance graph function as triggers to this process. 
The approach is implemented as the open-source Reuse and Adapt framework for Simulation Experiments (RASE). 

We demonstrate the utility of our framework in two simulation studies, from demography and cell biology.
We show that simulation experiments as well as other provenance entities can be effectively reused and exploited for automatically generating a simulation experiment.

The outline of this paper is as follows.
First, in Section~\ref{sec:background} we introduce the prerequisites for our approach, including provenance and means for explicitly specifying simulation experiments.
Then, we go deeper into when and which simulation experiments are typically reused (Section~\ref{sec:scenarios}).
In Section~\ref{sec:framework}, we present our reuse and adapt framework for simulation experiments.
This is then followed by implementation details in Section~\ref{sec:implementation}. 
In Section~\ref{sec:casestudies}, we apply our framework to two simulation studies from demography and cell biology, respectively: one aimed at developing a simulation model to study the impact of information flow on migration, the other aimed at revealing crucial mechanisms of a central signaling pathway.  
We finish the paper with related work in Section~\ref{sec:relatedwork}, and conclusions and future work in Section~\ref{sec:conclusions}.

\section{Background}
\label{sec:background}
Our approach is based on two main ingredients: 1) the concept of provenance for simulation studies and 2) explicitly specified simulation experiments. This section provides important background knowledge on these topics.

\subsection{Provenance of Simulation Studies}
\label{sec:provenance}
Provenance refers to gathering information about how a product has been generated. 
As stated by the W3C Provenance Working Group, provenance provides ``information about entities, activities, and people involved in producing a piece of data or thing, which can be used to form assessments about its quality, reliability, or trustworthiness''~\cite{Moreau2013provenance}. 
To apply provenance to products of modeling and simulation studies requires to identify the central activities and products of modeling and simulation, and to put those into relation in a directed acyclic graph~\cite{Ruscheinski2017provenance}.  
To record provenance information about simulation studies different possibilities exist. 
For example during conducting a simulation study, artifact-based workflows~\cite{Ruscheinski2019artifact} allow to collect fine-grained provenance information about the various activities, information sources, and products. 
Recording the information within a graph database enables zooming in and out of simulation studies and querying provenance information about simulation studies on demand~\cite{Ruscheinski2019capturing}. 
More coarse-grained provenance information, typically manually recorded after a simulation study has been conducted (ideally by those who conducted the simulation study), still reveals important information about the development process of individual simulation model and its potential for reuse, or of entire sets of simulation models~\cite{Budde2021relating}. 
In the latter case, the provenance information elucidates a family of simulation models with their specific relations, partly shared data and information sources, and close ties realized by cross validations.

To represent the provenance information of our simulation study we specialize the PROV Data Model (PROV-DM)~\cite{Moreau2013provenance}. 
We distinguish between entities and activities, and relate those by relations of being used or being generated. 
In an earlier specialization of PROV-DM for telling the tale behind a simulation model or a simulation study~\cite{Ruscheinski2018towards, Budde2021relating}, we identified important entities and activities of relevance in developing a simulation model or conducting a simulation study. 

\begin{figure}
	\centering
	\includegraphics[width=0.7\textwidth]{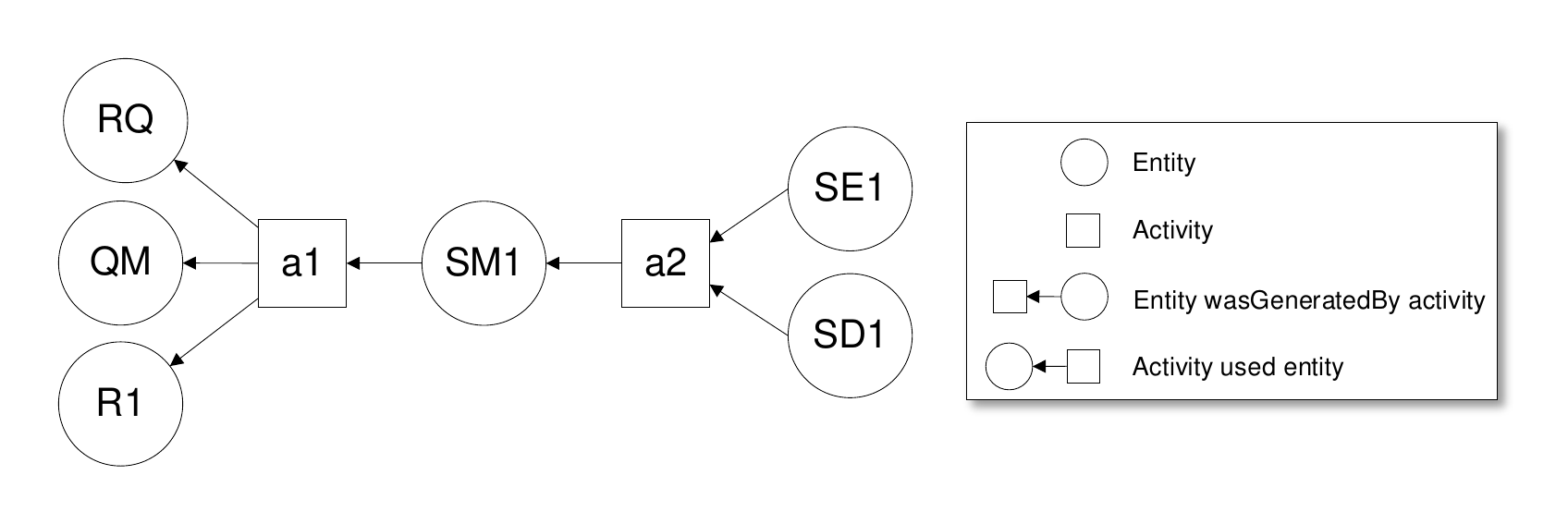}
	\caption{Example of a provenance graph in the PROV-DM notation~\protect\cite{Belhajjame2013prov}.}
	\label{fig:provenanceExample}
\end{figure}

In the following, we will distinguish as entities the research questions (RQ), simulation models (SM), simulation experiments (SE), simulation data (SD), other data (D), e.g., from the wet-lab or surveys, requirements (R), qualitative model (QM), assumptions (A), theories (T), and other information (O).  
The activities will represent the typical activities of the modeling and simulation life cycle such as building a simulation model, its calibration, or validation~\cite{Balci2012lifecycle, Sargent2013verification}. 
According to PROV-DM, activities and entities are related by the relations \emph{wasGeneratedBy} (activity $\leftarrow$ entity) and \emph{used} (entity $\leftarrow$  activity). 
Although these entities, activities and dependencies should allow documenting provenance for a large number of simulation studies, largely independently of the application domain or simulation approach, provenance models can easily be extended.
Overall, we are aiming at a high-level view of the simulation study (which can be created from every lower-level provenance graph by aggregating smaller (similar) activities and then using only the most recent entities as inputs and outputs to the aggregated activity).
For example, instead of representing multiple conceptual modeling and model building steps that consider the research question, qualitative model, and assumption one after the other, and producing various intermediate versions of the simulation model, the aggregated view would show simply a model building activity that takes as inputs the research question \texttt{RQ}, the qualitative model \texttt{QM}, and the requirement \texttt{R1}, and produces a simulation model (see activity a1 in Figure~\ref{fig:provenanceExample}).

\subsection{Explicit Simulation Experiments}
\label{sec:specifications}
The execution of simulation experiments plays a central role in the modeling and simulation life cycle. 
Treating simulation experiments as first class objects facilitates their generation, reuse, and repetition. 
This also enhances the reproducibility of modeling and simulation~\cite{Taylor2018crisis}.
Not surprisingly, reporting guidelines for simulation studies increasingly demand information about simulation experiments, e.g., the ``Strengthening the reporting of empirical simulation studies'' (STRESS) in the field of operational research and management sciences~\cite{Monks2019strengthening}, or the ``Overview, Design concepts and Details'' (ODD) protocol for agent-based models in ecology~\cite{Grimm2020odd}. 
Another strand of research focuses on specifying simulation experiments unambiguously in a machine accessible and executable manner.  
Especially in the context of discrete-event simulation and computational biology, a variety of approaches have been developed.
These include experimental frames~\cite{Zeigler1984multifacetted}, 
domain-specific languages and formats ~\cite{Waltemath2011reproducible,Perrone2012safe,Ewald2014sessl,Dayibas2019xperimenter}, and 
model-based approaches for experimental design~\cite{Teran2015model}.

Model-driven approaches are already a common technique for the development of simulation models~\cite{Cetinkaya2015model}.
In previous work, we developed metamodels for simulation experiments (also called experiment schemas) based on JSON to have an exchangeable format that allows reuse across tools and approaches~\cite{Wilsdorf2019simulation}.
Our metamodels define the various parts of a simulation experiment. 
Some parts are specific to the type of simulation, i.e., the initialization of the model, the choice and parametrization of the simulator, and setting up the observation intervals and measures.
Other parts are specific to the experiment type, such as parameter scans, local or global sensitivity analysis, statistical model checking, or optimization.
In Listing~\ref{lst:jsonExample} the full specification of a parameter scan is given.
First, the basic experiment is specified (lines 2--20) with a model named \texttt{sir.mlrj}, a stochastic simulator, 100 stochastic replications, stop time 80, and the model compartments \texttt{s}, \texttt{i} and \texttt{r} observed every 20 time units.
Then, the parameters scan (lines 21--26) is specified for the parameters \texttt{k1} and \texttt{k2}, providing the minimum and maximum values as well as a scanning interval for each parameter. 

\begin{figure}[h]
\begin{lstlisting}[caption=Experiment specification of a parameter scan for a SIR model. Specified using the JSON-based format by \protect\cite{Wilsdorf2019simulation}., label=lst:jsonExample]
{
	"model": {
		"modelPath": "./sir.mlrj"
	},
	"simulation": {
		"simulator": "SSA",
		"replications": 100,
		"stopCondition": {
			"stopTime": 80
		}
	},
	"observation": {
		"observables": {
			"observationExpression": ["count(\"s\")", "count(\"i\")", "count(\"r\")"],
			"observationAlias": ["susceptible", "infected", "recovered"]
		},
		"observationTime": {
			"observationTime": [0, 20, 40, 60, 80]
		}
	},
	"parameterScan": {
		"factorName": ["k1", "k2"],
		"factorMinimum": [0.5, 0.5],
		"factorMaximum": [2.0, 2.0],
		"interval": [0.1, 0.1]
	}
}
\end{lstlisting}
\end{figure}

\section{Typical Reuse Scenarios in Simulation Studies}
\label{sec:scenarios}
First, we want to describe some example scenarios to illustrate when our framework could be of assistance.
These are reuse scenarios that occur in simulation studies where the development of a valid simulation model is the focus. 
In these kinds of studies, the reuse of simulation experiments is typically evoked by the creation of a new simulation model (version).
Many scenarios refer to the reuse of simulation experiments within the same simulation study.
However, simulation experiments may also be reused across studies, i.e., the reused entities are taken from another related simulation study, presuming that the provenance graphs of the two studies are connected.

\subsection{Repeated Model Validation (Regression Testing)}
Validation is an important task in the modeling and simulation life cycle.
It is the process of substantiating whether the model behaves consistently with our expectations, e.g., regarding data which has been measured in the real system~\cite{Reinhardt2019valid}.
However, the model development is usually not completed after the first validation. 
Further model features are added, and the modeler moves again through the phases of the modeling and simulation life cycle.
Making changes to the simulation model then, of course, requires re-validating it~\cite{Peng2016reusing, Ruscheinski2019artifact}.
In software development, successive validation is known as regression testing: ``Regression testing is performed between two different versions of software in order to provide confidence that the newly introduced features of the [system under test] do not interfere with the existing features''~\cite{Yoo2012regression}.

A special case of this scenario for repeating validation experiments, is model composition.
Frequently, models are not built from scratch but are created by composition of existing models.
The composed models may have been developed as separate modules during the same simulation study, or may originate from different, previously conducted simulations studies.
Once two or more models have been merged, it should be evaluated if everything still works as expected~\cite{Peng2017reusing}.
Thus, validation experiments that were conducted with the individual models are typically repeated with the composed model.

\subsection{Repeated Model Calibration}
Calibration, also called model fitting, is the process of finding a parameterization of the model which can reproduce observed behavior of the real system~\cite{Reinhardt2019valid}.
While calibration and validation both typically relate to real data, the conclusion drawn from them are essentially different.
Whereas calibration refers to adjusting the input parameters such that the resulting agreement of the model output with a chosen set of experimental data is maximized, the goal of validation is to establish confidence in the model predictions~\cite{Trucano2006calibration}.
Therefore, if previous calibration experiments exist in a simulation study, they should be repeated before the validation experiments.
Consequently, when a new model version is produced, first calibration experiments are repeated. 
Only if the calibration was successful, can the validation be attempted.
If the model cannot be calibrated such that it reproduces the data with sufficient accuracy, further model revisions are necessary.

\subsection{Repeated Model Analysis}
In simulation studies, many experiments are conducted that do not serve as validation or calibration.
Nevertheless, they still reveal important information about the model, e.g., via parameter scans, optimization, sensitivity analysis, perturbation analysis, or time course analysis~\cite{Budde2021relating}.
It could be argued that all experiments contribute to the validation of the simulation model as they increase our trust that the right model has been built. 
In the context of this paper we assume validation experiments to be experiments distinguished as such by the modelers, and whose outcome can be evaluated as a \emph{success} or \emph{failure}, see also~\cite{Ruscheinski2019artifact}.
Especially sensitivity analysis is increasingly becoming an integral part of simulation studies as it allows quantifying how the parameters contribute to the uncertainty in the model output \cite{Coveney2021ReliabilityAR}.
It is becoming good practice, to attach each model version with uncertainty and sensitivity information. 
Therefore, automatically reusing and repeating sensitivity analysis experiments after each major model version is  instrumental in enhancing the quality of simulation models.

\subsection{Cross-Validation with Related Simulation Studies}
Cross-validation (or model alignment~\cite{Axtell1996aligning}) is the process of comparing a simulation model with another, independently developed model. 
In particular, models that deal with a similar research question should be able to reproduce each other's results.
By comparing (and i.e., validating) simulation models with other, already calibrated and validated models, a ``domain validity'' can be achieved~\cite{Axtell1996aligning} and overall confidence in the models can be established.
To facilitate cross-validation, simulation experiments conducted with related, validated models should be reused and adapted in the validation activities of the current simulation study.

\subsection{Comparison of Alternative Implementations}
Usually, there is not just one way of modeling a system.
From the same conceptual model different computerized models can be built.
For example, the same model can be simulated using discrete event simulation (based on Continuous-time Markov Chains) or using System Dynamics (based on Ordinary Differential Equations ODEs), and with or without spatial features.
Comparing alternative modeling and simulation approaches is crucial, e.g., to uncover discrepancies in simulation results, or to find a more efficient implementation in terms of simulation runtime.

The comparison is done by reusing and reapplying simulation experiments that were conducted with the other model implementation. 
Sometimes this comparison is done as part of the same simulation study, e.g., when using the ``modeling in pairs'' technique, which shall prevent modelers from making bad implementation-related choices too early in the model development process~\cite{Reinhardt2019developing}.
However, the re-implementation and thus improvement of a previous model could also be the primary goal of a simulation study.  
In this case, simulation experiments have to be reused across simulation studies.

\section{Reuse and Adapt Framework for Simulation Experiments}
\label{sec:framework}
To automate the reuse of simulation experiments like the scenarios described above, we develop a Reuse and Adapt framework for Simulation Experiments (RASE).
The framework presents a provenance-based mechanism for reusing simulation experiments either within or across simulation studies.  
It exploits the observation that certain activities of the modeling and simulation life cycle produce characteristic patterns in the provenance graphs. 
Based on such patterns, the production rules of a graph transformation system can be constructed. 
In the following, we first identify and characterize these provenance patterns.
Next, we define a provenance graph transformation system for producing new experiment activities based on the last modeling activity and previously conducted simulation experiments.
Finally, we describe how to adapt simulation experiment specifications to the context information from the new simulation model, which is also given by provenance.

\subsection{Provenance Patterns in Simulation Studies}
Provenance patterns are the central building block for our approach.
Associating provenance patterns with a semantics enables us to interpret what happened during the simulation study.
In particular, patterns are used in three ways by our approach.
\begin{enumerate}
	\item \emph{Trigger patterns} initiate the reuse, adaptation and execution of certain experiments. 
	For us, trigger patterns always denote activities that produce a simulation model. 
	Note that, especially if more fine-grained provenance information is available, other triggers that do not produce a simulation model might be possible.
	\item  \emph{Experiment patterns} are used to identify and retrieve previous simulation experiments. They describe either an experimentation activity with a specific role, e.g., experiments used for calibration, validation, or analysis~\cite{Budde2021relating}, or experiments of a specific type, such as sensitivity analysis or statistical model checking. 
	\item Eventually, from the provenance information a new activity including the adapted simulation experiment will be generated. 
	Therefore, patterns are used as \emph{blueprints} for creating a new activity and connecting all entities correctly.
\end{enumerate}

\begin{figure}
	\centering
	\includegraphics[width=\textwidth]{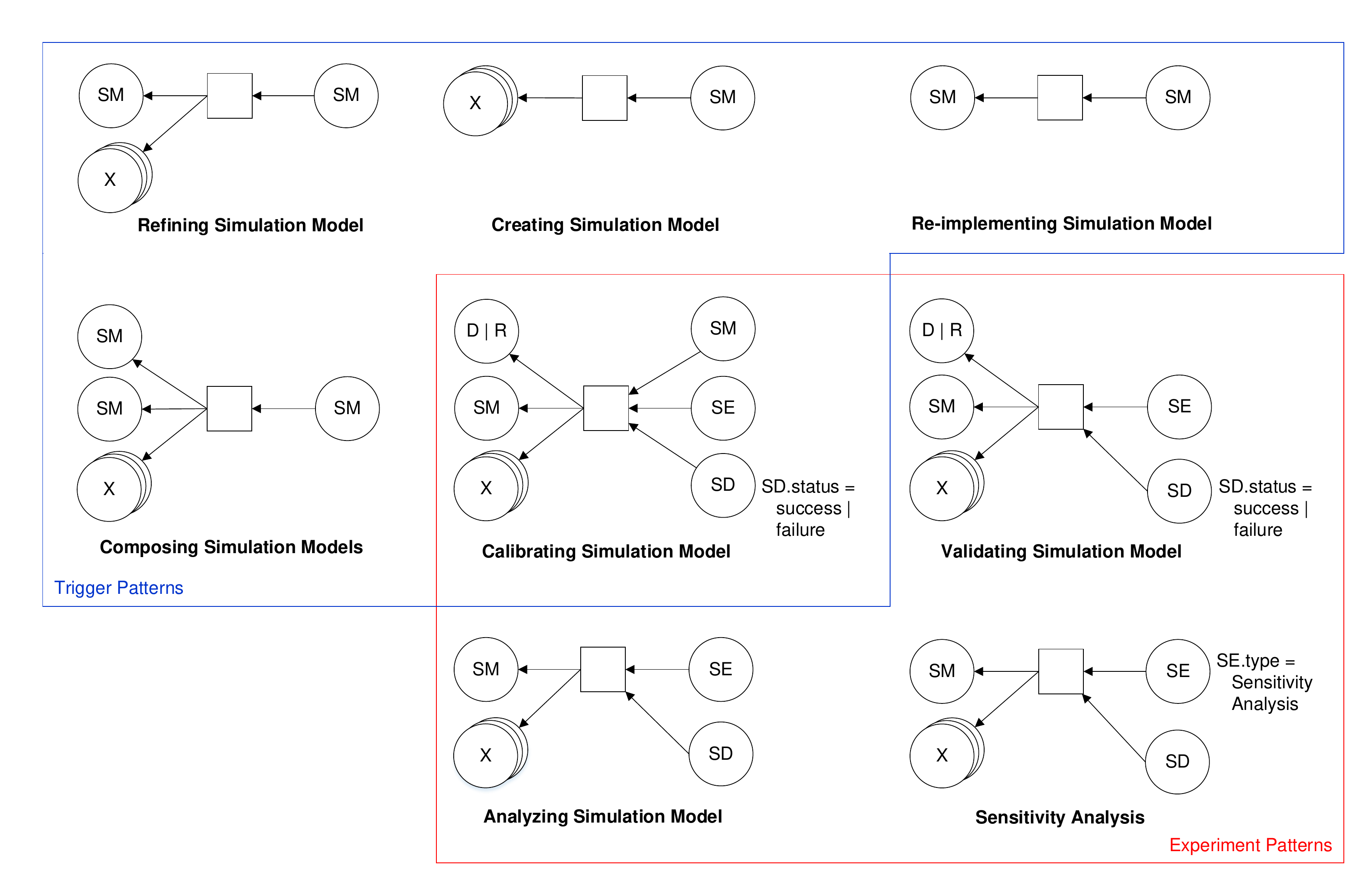}
	\caption{Provenance patterns for interpreting changes in the simulation models (trigger patterns), and finding suitable simulation experiments to reuse (experiment patterns). The patterns show what entities are required as inputs and outputs for a certain type of activity. The multi-entities (X) are used to capture all remaining inputs, however, they must not contain entities of type simulation model (SM).}
	\label{fig:expPatterns}
\end{figure}

In the following, we predefine a number of provenance patterns that are necessary for supporting the scenarios described in Section~\ref{sec:scenarios}.
The patterns are illustrated in Figure~\ref{fig:expPatterns}.

\paragraph{``Refining Simulation Model'' Pattern:}
Model refinement or extension is the typical model building step during a simulation study.
It involves a single simulation model, and produces a new one.
Usually, a variety of other entities are used during a model building step such as research questions, qualitative models, theories, assumptions, requirements, and data.  
If the used model entity belongs to a different simulation study than the generated model, the pattern describes the extension or refinement of an existing typically already validated model.

\paragraph{``Creating Simulation Model'' Pattern:}
When a simulation model is created from scratch no simulation model is used. 
However, analogously to the refining pattern, various entities can be used: research questions, qualitative models, theories, assumptions, requirements, data, or other information sources.
The output of the ``creating simulation model'' activity is an initial simulation model that is either entirely new, or just has not been linked to another study yet.

\paragraph{``Re-Implementing Simulation Model'' Pattern:}
Sometimes models are not refined or extended but re-implemented using other tools or languages.
When re-implementing a simulation model, the provenance graph reflects this as a model building activity, i.e., a simulation model is used, and another simulation model is generated by the activity.
But in contrast to extending or refining, no new conceptual materials (such as a qualitative model, or input data) are used.

\paragraph{``Composing Simulation Model'' Pattern:}
The composition pattern involves two simulation models as input to an activity (used-dependency).
These are fused into a composed model, i.e., the output of this activity.
Similarly to the refining or creating simulation model pattern, further entities can be used that deliver context information about why and how to combine the models.
Moreover, the used models and the composed model may belong to different simulation studies.

\paragraph{``Calibrating Simulation Model'' Pattern:}
In a calibration experiment the model parameters are fitted with the help of additional context data.  
Thus, the input of the calibrating simulation model pattern consists of a simulation model as well as data or alternatively a requirement entity (if the requirement, e.g, expresses properties of the target trajectory and a distance measure). 
The output of a calibration activity is a simulation experiment containing the experiment specification, a data entity containing the simulation output and the calibration status (success/failure), as well as a modified (fitted) simulation model.
Of course, further entities are allowed as input to the calibration, e.g., assumptions or theories or input data, but these are not required.
Also simulation experiments may be used as input to a calibration.
This indicates that the new experiment was created by reusing a previous experiment.

\paragraph{``Validating Simulation Model'' Pattern:}
The validating simulation model pattern looks in large parts similar to the calibration patterns.
Here, also either a data or requirement entity is required.
As in the case of calibration, other entities can be used, e.g., to input data containing the initial configuration, or reused simulation experiment. 
Products of the validation activity are a simulation experiment and simulation data indicating success or failure of the validation.
In contrast to calibration, no new simulation model is generated.

\paragraph{``Analyzing Simulation Model'' Pattern:}
An analyzing simulation model activity uses a single simulation model as input, produces a simulation experiment and simulation data.
In contrast to calibration and validation activities, they do not require specific input entities and their results are not interpreted as success or failure.

\paragraph{``Sensitivity Analysis'' Pattern:}
Especially with analysis experiments, we could also ask more specifically about an experiment type that was conducted during the simulation study.
Therefore, the experiment patterns can be refined by including meta-data from the information model of the experiment entity.
Thereby, for example the analysis pattern can be refined to a ``sensitivity analysis'' pattern.
This, of course, assumes that the experiment type was made explicit as an attribute in the information model of the experiment entity.

\subsection{Experiment Reuse by Provenance Graph Transformation Rules}

The reuse, adaptation, and generation of simulation experiments will be based on a graph transformation system.
The graph transformation system is given by a set of production rules $R$, which we will also call the reuse rules.
A reuse rule is given in the form $(T, E, C) \rightarrow E_{gen}$.
The left-hand side of the rule consists of a trigger pattern $T$ (a pattern that generates a simulation model), an experiment pattern $E$ (a pattern that generates a simulation experiment), and a condition $C$ relating the two.
The condition is a Boolean function that evaluates on the trigger and the experiment pattern.
For this, various helper predicates (e.g., \textit{isBasedOn(Simulation Model, Simulation Model)} or \textit{isValidated(Simulation Model)}) are used, which also are small provenance patterns.
The right-hand side of a rule takes the entities from the left-hand side and describes how to extend a given provenance graph $G$.
Note that, we do not allow rules that modify or delete existing nodes, thus, it suffices that the right-hand side specifies the generation instructions $E_{gen}$, i.e., what the new activity and entities should be and how to connect them to the entities of the trigger and experiment patterns.
The generation part $E_{gen}$ should match one of the experiment patterns. 
This way it can be recognized and reused in future model development cycles.

\begin{figure}
	\centering
	\includegraphics[width=\textwidth]{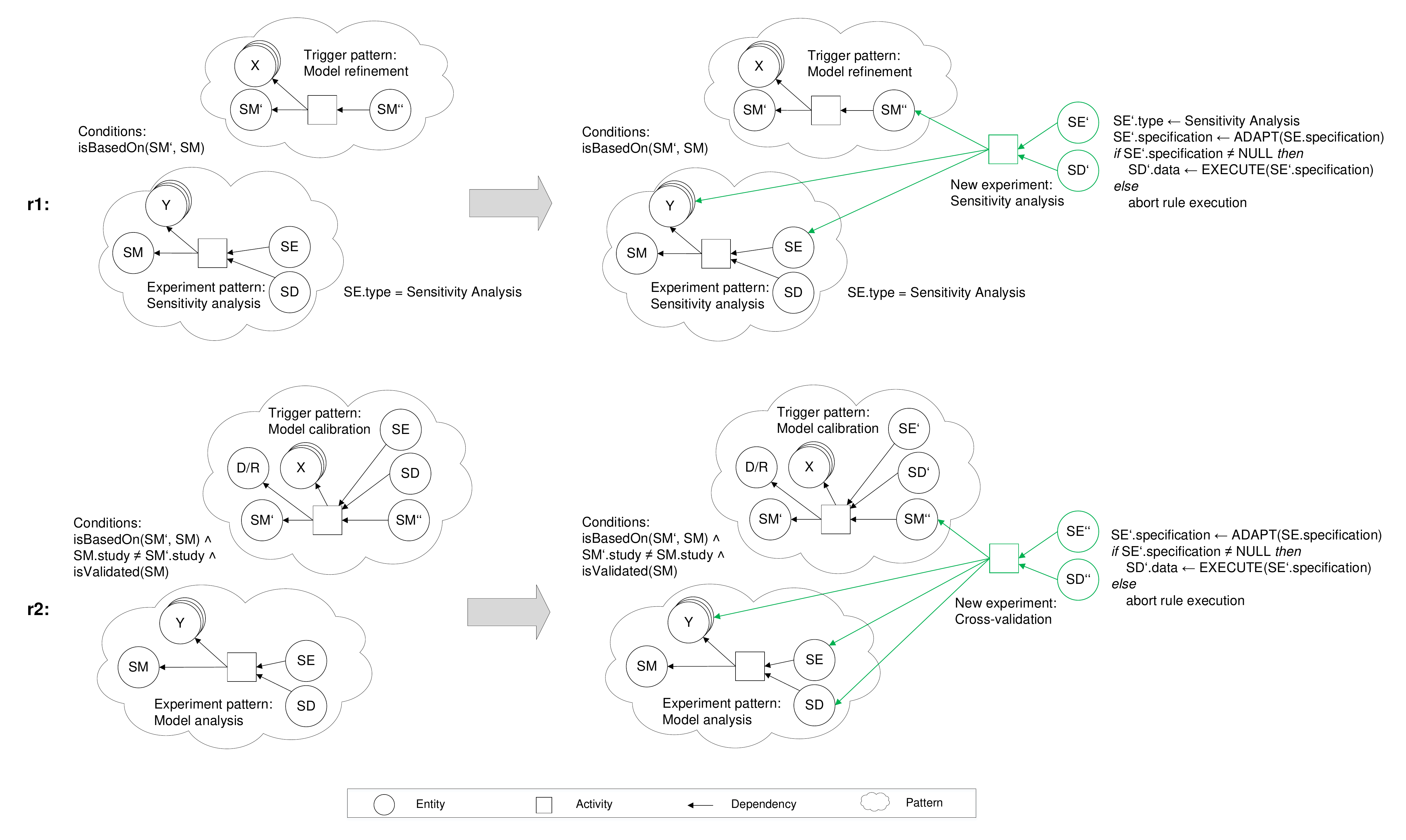}
	\caption{Rule r1 describes the repetition of a sensitivity analysis. Rule r2 shows the cross-validation scenario. In both rules, the left-hand side specifies a trigger pattern, a pattern of a previous simulation experiment, and additional conditions. The right-hand side extends the provenance graph by a new experiment activity (shown in green) and fills the information models of the generated entities using the functions ADAPT and EXECUTE.}
	\label{fig:ruleExamples}
\end{figure}

For our framework, we predefine a set of generation rules which map to the scenarios from Section~\ref{sec:scenarios}.
The rule set can be customized by enabling or disabling rules depending on the user's level of expertise.
Furthermore, custom rules can be created for specific settings by combining trigger patterns and experiment patterns as needed.
Figure~\ref{fig:ruleExamples} shows two exemplary rules.

The first rule describes the generation of a sensitivity analysis. 
It takes a model refinement step as a trigger ($T$), and the sensitivity analysis pattern for finding previous experiments ($E$). 
As condition ($C$) for searching in the provenance graph, the simulation model used in the refinement has to be based on the model used in the sensitivity analysis. 
From this information, a new activity is generated that takes the latest simulation model (from the trigger pattern), as well as other inputs from the previous experiment activity  (accumulated in the multi-entity Y). 
To denote that an experiment was reused during this step, also the old simulation experiments SE is taken as input.
The output of the new activity is a simulation experiment entity with type sensitivity analysis and an adapted experiment specification. 
Moreover, a simulation data entity is generated.
It is filled with the results from the experiment execution of the adapted experiment specification.
If the old specification could not be adapted to fit the new model, the experiment generation (i.e., the execution of the current rule) is aborted. 
This might be the case if important species or parameters of the old model do not exist anymore.

The second rule example presents the cross-validation scenario.
Here, model calibration is used as a trigger activity. 
Suitable experiments to reuse are identified using the analysis pattern.
However, further conditions have to apply, i.e., the simulation model used in the trigger pattern has to be based on the simulation model used in the analysis pattern, they have to belong to different studies, and the model used in the analysis has to be a validated.
The generated activity corresponds to the validation pattern.
In contrast to the generation part in r1, here the output data of the previous experiment is taken as input to allow for comparison of the two models.

The rule matching procedure is given in Algorithm~\ref{alg:reuse}.
It is started each time a new activity $a$ including dependencies and newly generated entities was completed and added to the provenance graph $G$.  
Note that, we do not need to evaluate the rules in a specific order.
Experiments on the same simulation model can be executed in parallel, as their results are independent from one another. 
However, if a rule produces a new simulation model (i.e., if the generation pattern is a model calibration), a new round of rule matching may be triggered.
Depending on the defined rule set, the next experiments triggered by this calibration might depend on the results of the experiments executed in the previous round.
Therefore, it is necessary to first wait until all experiments executed with the previous model are completed. 

Then, for each rule the trigger pattern is matched in $G$.
Here the most recent activity $a$ serves as anchor point, i.e., the trigger pattern can only match at that activity (identified by its $id$).
This is used to limit the search space and to only retrieve subgraphs relevant to the current activity and thus the current simulation model.
If, for example, already a cascade of generated simulation experiments exists, not all of these need to be reused, but we only use the latest version conducted with the predecessor model version.
Note that according to our definitions, a trigger pattern only contains one activity, which can be accessed by $T.activity$.

If the trigger of the rule matches at the current activity, all occurrences of $E$ are collected that fulfill the conditions in relation to the matched trigger.
We will often find multiple reusable experiments, for example, various validation experiments where each checks a different behavioral property of the model given by a requirement entity.
For each of these matches $e \in M$ then the right-hand side of the rule is executed, and the generation pattern $E_{gen}$ is collected within a new graph ($N$).
The execution of the rules does not only create the graph structure defined by $E_{gen}$, but also fills the information models of the new entities by applying the functions ADAPT and EXECUTE (see example rules in Figure~\ref{fig:ruleExamples}). 
ADAPT takes the old experiment specification and generates an adapted version for the new simulation model.
The function EXECUTE then takes the new experiment specification, runs it, and collects the simulation data.
Subsection~\ref{sec:adaptations} describes these steps in detail. 

When all rules have been evaluated, these new experiments can be added all at once to the provenance graph, i.e., $G = G + N$.

\begin{algorithm}[t]
	\DontPrintSemicolon
	\vspace{0.5ex}
	\KwInput{a provenance graph $G$,
		a set of reuse rules $R$,
		the current activity $a$}
	\vspace{0.5ex}
	wait for all experiments executed with previous model to complete \\
	$N \leftarrow \emptyset$ \\
	\ForEach(){$(T, E, C) \rightarrow E_{gen}$ \textbf{in} $R$}{
		$t$ $\leftarrow$ find match of $T$ in $G$ where $T.activity.id = a.id$\\
		\If(){$t \neq \textsc{null}$}{
			$M$ $\leftarrow$ find all matches $e$ of $E$ in $G$ where $C(t,e) = true$ \\
			\ForEach(){$e$ \textbf{in} $M$}{
				$N \leftarrow N + E_{gen}$
			}
		}
	}
	$G \leftarrow G + N$
	\caption{Rule matching at activity $a$}
	\label{alg:reuse}
\end{algorithm}

\subsection{Adaptation, Generation and Execution of the Experiment Specifications} 
\label{sec:adaptations}

The old experiment specifications extracted via the experiment patterns may not be executable with the current model (version).
To evaluate whether changes have to be made to the experiment specification, the contexts of the old and new simulation experiments need to be taken into account.
Under the term context, we subsume all entities that participated in the model building process.
These may be, e.g., the qualitative model, assumptions, data and information sources, research question, assumptions, or theories, which are also known as the conceptual model~\cite{Robinson2015conceptual}.
We trace these entities in the provenance graph and check newer versions (i.e., since the last experiment execution) for relevant information.

\begin{figure}
	\centering
	\includegraphics[width=\textwidth]{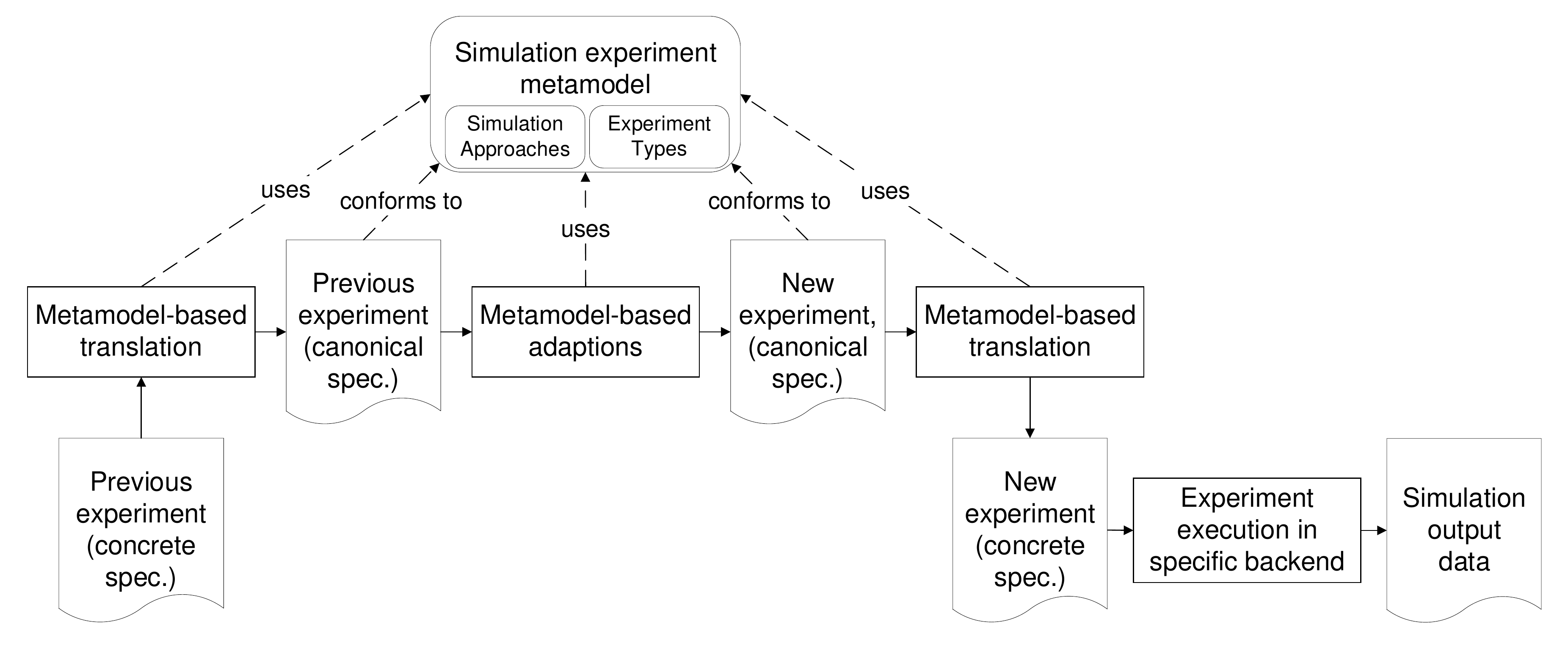}
	\caption{Adapting and generating simulation experiments based on metamodels. The original pipeline and the metamodels were defined in~\protect\cite{Wilsdorf2019simulation}.}
	\label{fig:generationPipeline}
\end{figure}

But before any adaptations can be made, we have to make all the parts of the experiment specifications easily accessible.
Therefore, they are translated to an intermediate representation, which we will call the canonical form. 
Here we use the JSON-based format introduced in Section~\ref{sec:specifications}.
In the canonical form, each part of the experiment specification is assigned a quasi-standardized term.
Based on these terms, adaptions can be defined just once for a variety of different modeling and simulation tools and approaches.
For the translation, we use a metamodel of simulation experiments, see Figure~\ref{fig:generationPipeline}.
The metamodel defines the vocabularies for various simulation approach, e.g., discrete-event simulation.
Experiment specifications given, e.g., in SESSL, SED-ML, or Python, are assigned terms such as \texttt{modelPath}, \texttt{stopTime}, or \texttt{replications}.
Moreover, the metamodel provides experiment type-specific vocabulary, e.g., \texttt{parameterDistribution} for sensitivity analysis, or \texttt{propertyExpression} for statistical model checking.

Which of these parts needs to be adapted in the selected experiment specifications can be derived from the information models of the provenance entities.
The better these are structured and substantiated with formal expressions, the better they can be exploited automatically.
Plenty of work on conceptual modeling~\cite{Robinson2015conceptual}, workflows~\cite{Ruscheinski2019artifact}, and provenance ontologies~\cite{Budde2021relating} focuses on the contents of the various entities.
In the following we list some adaptations that are currently supported in our framework:
\begin{itemize}
	\item The current simulation model may be located in a different folder than the old model, or it may have a different name. 
	To update the experiment specification accordingly (i.e., the \texttt{modelPath}), meta-information from the simulation model entity can be used. 
	\item Parameter names and initial values belong to every experiment specification. 
	If the name of a parameter changed or if the initial values of model were revised, they also need to be updated in the reused experiment specification (i.e., \texttt{parameterName}, \texttt{parameterValue}). 
	The qualitative model is the entity that comprises valuable information about model inputs.
	Also the simulation model itself might be annotated with useful information.
	\item Similarly to the set of parameters, the observed species used in the \texttt{observationExpression} of the experiment specification might have to be adapted. 
	Again the qualitative model or the simulation model can provide information about the model outputs.
	\item For simulation experiments such as parameter scans or sensitivity analyses, a central aspect of the specification is the experiment design. 
	It includes specifying the \texttt{minimumValue} and \texttt{maximumValue} as well as the \texttt{distribution} of each factor (i.e., parameter varied during the experiment). 
	Information about the values of a factor may be available in the form of assumption entities.
	\item Other experiment types, such as statistical model checking, rely on formally specified requirements. 
	Occasionally, the \texttt{propertyExpression} may have changed since the last execution of the simulation experiment, which will be indicated by an update on a requirement entity in the provenance graph.
	If the formula is included explicitly in the requirement entity, it can automatically be inserted into the new experiment specification. 
\end{itemize}

While some changes are easy to detect, others require special mechanisms for comparing provenance entities.
One option is annotating the information models with tags from various ontologies.
These can be ontologies to identify a parameter or compartment (e.g., the Gene Ontology (GO)~\cite{GeneOntology2018}) or ontologies to identify an algorithm or methodology (e.g., the Kinetic Simulation Algorithm Ontology (KiSAO)~\cite{Courtot2011controlled}, or an Ontology for Discrete-event MOdeling and simulation (DeMO)~\cite{Silver2011demo}).
Sometimes, if not enough meta-data is available in other entities, the simulation model specifications need to be compared.
In that case, BiVeS could be used for characterizing changes in SED-ML files~\cite{Scharm2015algorithm}.

After adapting the experiment specification in the canonical form, it translated back to a specification in a concrete language using the metamodel. 
Then, this concrete experiment specification is automatically executed in a corresponding modeling and simulation tool called backend (as illustrated in Figure~\ref{fig:generationPipeline}).

Backends are made available to the pipeline via bindings.
Which backend (and thus experiment language) is used, depends on the models and experiments at hand.
If experiments are reused within the same simulation study, typically the same language and backend can be used as with the original experiment.
On the contrary, when an experiment is reused across simulation studies, it is likely that these studies use different tools, and thus the new simulation experiment will be generated in a different language.
Information about which backend the current simulation study uses, can be determined by looking at other simulation experiments from that study.
While it is favorable to use tools consistently during a simulation study, the metamodel-based approach in general allows us to use any backend, as long as the kind of simulation model and the type of experiment are supported.
For example, a parameter scan of a model formulated in SBML~\cite{Hucka2003systems} could be conducted in both COPASI~\cite{Hoops2006copasi} and a general purpose programming language like Python or Java using the library LibSBML~\cite{Bornstein2008libsbml}.

Finally, after the experiment finished, the output data of the simulation is collected and attached to a new data entity.

Moreover, some simulation experiments also produce a new simulation model, and thus a simulation model entity needs to be created.
If the experiment automatically generates a new model specification, as with certain calibration procedures~\cite{Eckhardt2005automatic, Asadi2019building}, the model entity can be filled automatically.
Otherwise, we unfortunately rely on the user to modify the model specification manually based on the experiment results.

\section{Implementation and API}
\label{sec:implementation}
\begin{figure}
	\centering
	\includegraphics[width=\textwidth]{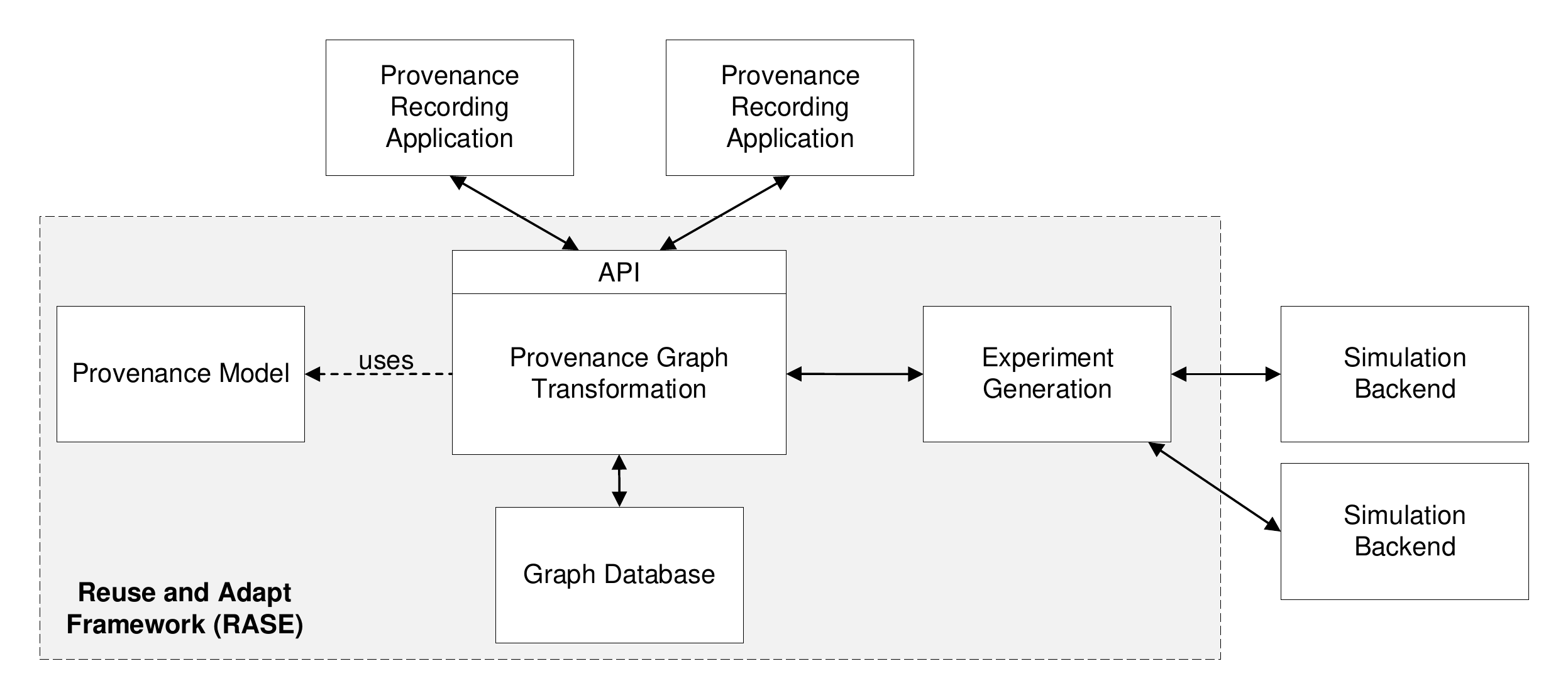}
	\caption{Overview of the reuse and adapt framework for simulation experiments.}
	\label{fig:provArchitecture}
\end{figure}

The implementation of our Reuse and Adapt framework for Simulation Experiments is available in a Git repository\footnote{\url{https://git.informatik.uni-rostock.de/mosi/exp-generation}}.
Figure~\ref{fig:provArchitecture} provides an overview of the software architecture, which was implemented using Java 1.8.
At the center is the Provenance Graph Transformation component which handles the reuse rules, manages access to the Graph Database, and coordinates with the Experiment Generation component.
An API is provided that allows connecting the framework to any application that produces provenance according to the Provenance Model introduced in Section~\ref{sec:provenance}, which is implemented in a separate module. 
The provenance recording applications may be stand-alone GUIs or workflow systems that capture provenance while the users run through a number of workflow stages. 
Furthermore, the API allows customizing the set of reuse rules and enabling graphical support depending on the user's needs.

Underlying our provenance graph transformation system is a Neo4j graph database~\cite{Miller2013graph}. 
Thus, the queries for the various provenance patterns are expressed in the Cypher query language~\cite{Francis2018cypher}.

As discussed earlier, we facilitate a previously developed toolchain~\cite{Wilsdorf2019simulation} for generating and executing simulation experiments. 
First, experiment specifications are transformed into an intermediate representation (JSON~\cite{Json}) based on metamodels (JSON Schema~\cite{JsonSchema}). 
Metamodels present a well-defined vocabulary and structure for different simulation approaches (such as discrete-event simulation) and different experiment types (such as sensitivity analysis or statistical model checking).
Metamodels are therefore valuable for identifying and accessing the parts of experiment specifications that have to be adapted based on provenance information. 
After the adaptations, experiment code is generated for the specific modeling and simulation tool, the experiment is executed, and the simulation data is collected.
The experiment generation toolchain already provides a number of bindings to tools, e.g., for SESSL~\cite{Ewald2014sessl} and ML-Rules~\cite{Helms2017semantics} (used in the Wnt signaling case study).
For the migration case study of this paper, we implemented an additional binding for simulation experiments conducted with Julia~\cite{Bezanson2017julia} and R~\cite{R-language}.

Note, that for this paper we focus on fully automated cases, i.e., everything from recording provenance to executing the generated experiment is done automatically.
However, in some cases it might be necessary to interact with the user due to the fact that the modeler still carries implicit knowledge about the simulation study. 
For example, the modeler may be asked to specify a requirement in a formal manner, or to review the generated experiment design.
For this purpose, the experiment generator includes a graphical user interface (GUI) based on JavaFX for presenting the generated experiment specifications to the user and to allow them to make changes~\cite{Wilsdorf2019simulation}.

\section{Case Studies}
\label{sec:casestudies}
We demonstrate the usefulness of our approach in two different simulation studies.
In the first case study, we will show a repeated sensitivity analysis of a model of migration routes.
In the second case study, we will reuse an experiment across simulation studies for cross-validation. 
Afterward, we discuss the advantages of our approach in the fields of sociology and cell biology.

Since in this paper we can only show a selected set of rules and patterns at work, with our source code we provide an additional, abstract simulation study to illustrate further reuse cases.
The abstract simulation study, inspired by epidemiological models, contains all the reuse cases described in the concept section, and does contain more adaptations of the experiment specifications, e.g., based on assumptions and qualitative models.

\subsection{Repeated Analysis of a Migration Model}

\begin{figure}
	\centering
	\includegraphics[width=\textwidth]{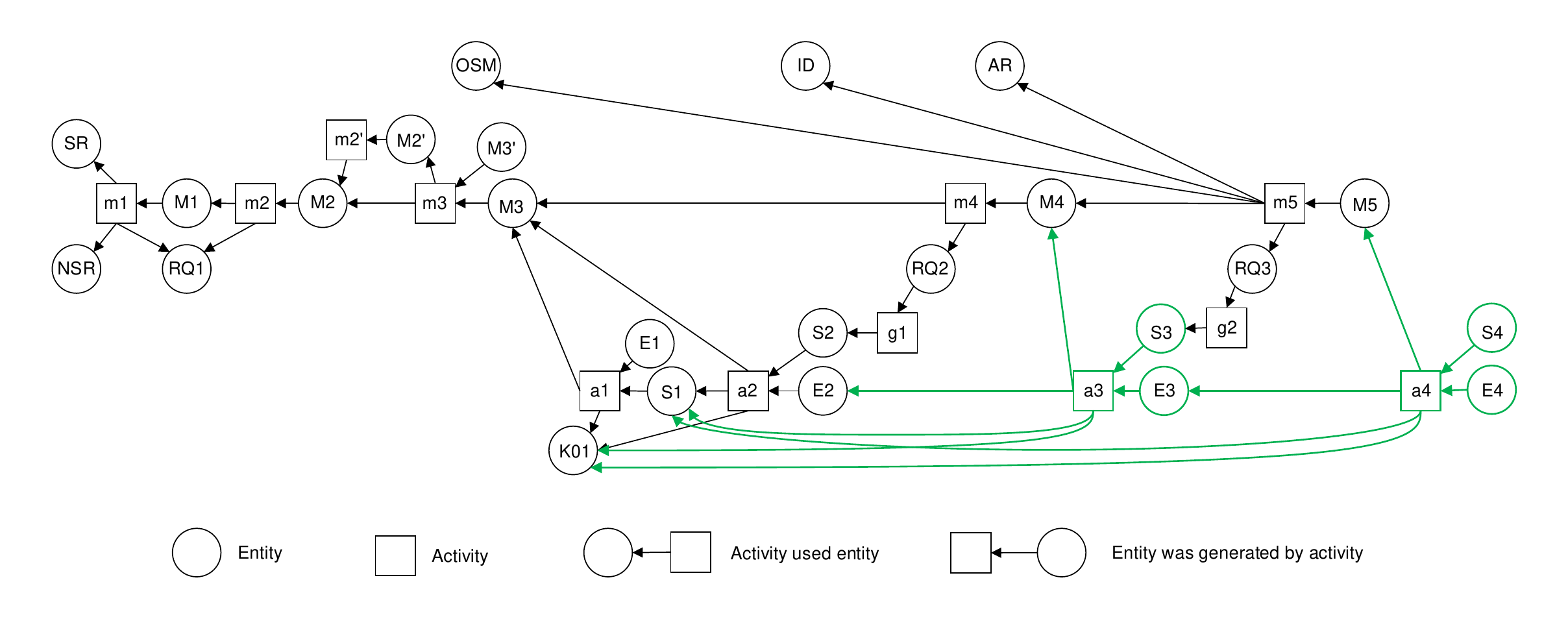}
	\caption{Provenance graph of the migration study published in~\protect\cite{Bijak2021book}. Activities, entities, and dependencies shown in green are generated automatically by our approach.}
	\label{fig:migrationGraph}
\end{figure}

The first case study refers to an agent-based model of asylum migration to Europe~\cite{Bijak2021book} focusing on the formation of routes and the propagation of rumors.
It aims to connect individual decisions (information transfer) on the micro-level to processes observed at the macro level (variability and optimality of migration
routes). 
Models are successively refined and analyzed\footnote{M1--M2: \url{https://github.com/mhinsch/RoutesRumours}, M3--M5: \url{https://github.com/mhinsch/rgct_data}} in response to developing theoretical lines of inquiry and the introduction of different data sources, which presents various opportunities for our framework to automatically reuse and generate simulation experiments.
The provenance of the study was described in~\cite{Bijak2021book}.
Here, we only show the parts that focus on the modeling and analysis activities (Figure~\ref{fig:migrationGraph}). 
Everything concerning psychological experiments and data processing is omitted.
Furthermore, we modified the provenance slightly to make the experiment specifications explicit ($E1-E4$).

The simulation study begins with the creation of a model M1, which already includes agent knowledge, social networks and information exchange with discrete-time behavior,
and model M2 which alters the simulated world from a grid-based layout to a more general and less densely connected graph topography~\cite{Bijak2021book}. 

During the model building steps m2' and m3, different versions of the originally time-stepped model (M2) are created to obtain more realistic time courses.
Model M3 then presents a continuous-time version of M2, implemented in Julia~\cite{Bezanson2017julia}.
With this continuous-time model M3, now first experiments are conducted.

In a1, a parameter scan is employed to reduce the 17 parameters to the 6 most influential factors.
Then, in a2 a sensitivity analysis is conducted on the selected parameters.
The analysis uses a Latin Hypercube sample to build Gaussian process emulators and carry out the uncertainty and sensitivity analysis.
In the original publication~\cite{Bijak2021book}, the analysis was conducted with the (GUI-based) tool GEM-SA.
For this case study, we prepared the same experiment as R scripts to have the experiment specifications accessible to our approach~\footnote{\url{https://github.com/jasonhilton/screen_run}}.

Following the experiment, the next model version is developed to include additional empirical data on information exchange and risk behavior, i.e., RQ2 and RF are used during the model refinement step m4.
This model refinement now triggers the rule matching of our approach.
Rule $r1$ from Figure~\ref{fig:ruleExamples} can be applied at activity m4 and the following match is found:
\begin{align*}
	match_1 = \{ &SM' \leftarrow M3, X \leftarrow \{RQ2, RF\}, SM'' \leftarrow M4, \\
	&SM \leftarrow M3, Y \leftarrow \{K01, S1\}, SE \leftarrow E2, SD \leftarrow S2
	\},
\end{align*}
where the left-hand side of the arrows represent the variable names given in the rules, and the right-hand side represent the names of the matched entities from the provenance graph in Figure~\ref{fig:migrationGraph}.

\begin{figure}[t]
	\begin{subfigure}[t]{0.48\textwidth}
		\includegraphics[width=\textwidth]{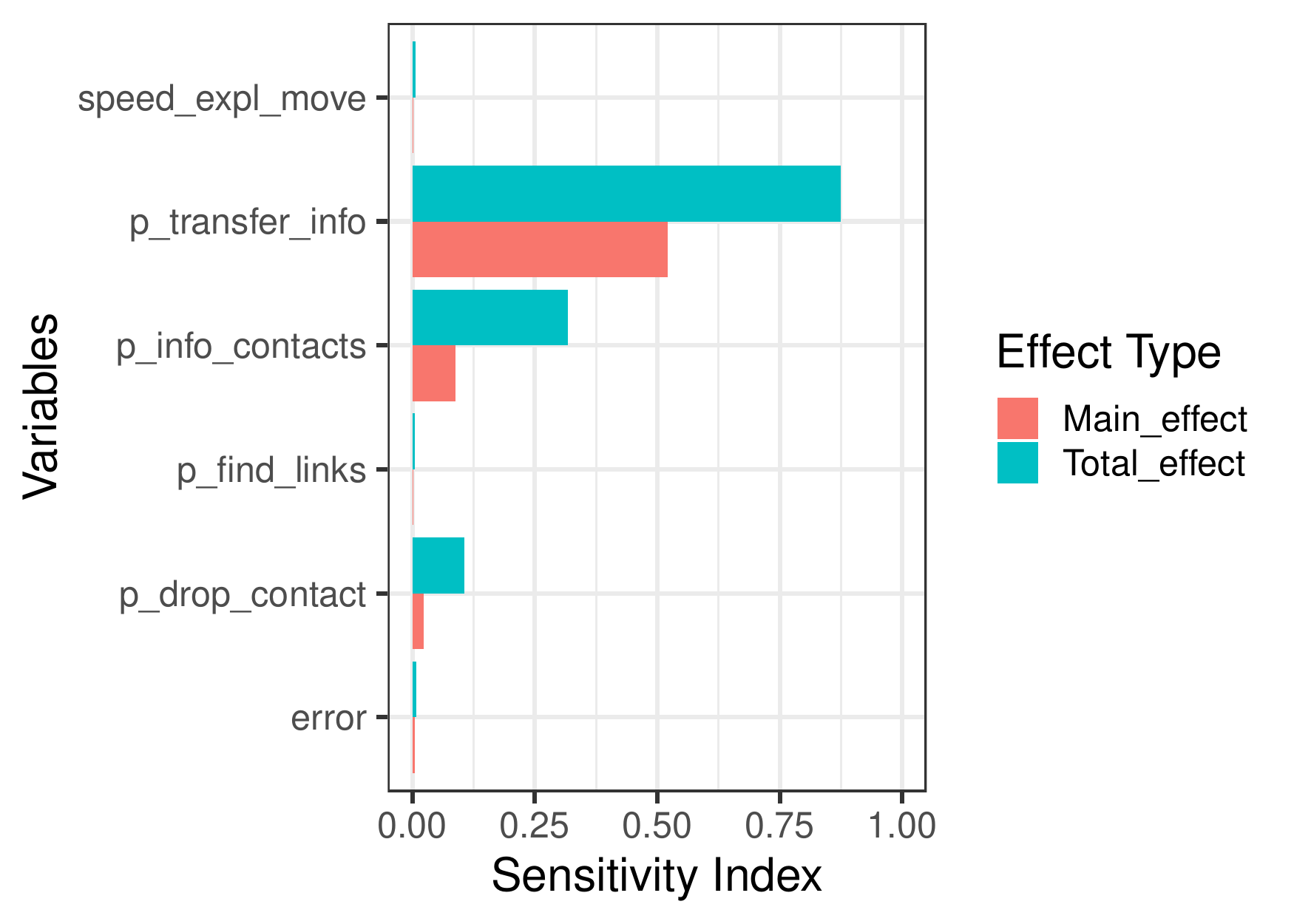}
		\caption{Main and total effects for M3}
	\end{subfigure}
	\hfill
	\begin{subfigure}[t]{0.48\textwidth}
		\includegraphics[width=\textwidth]{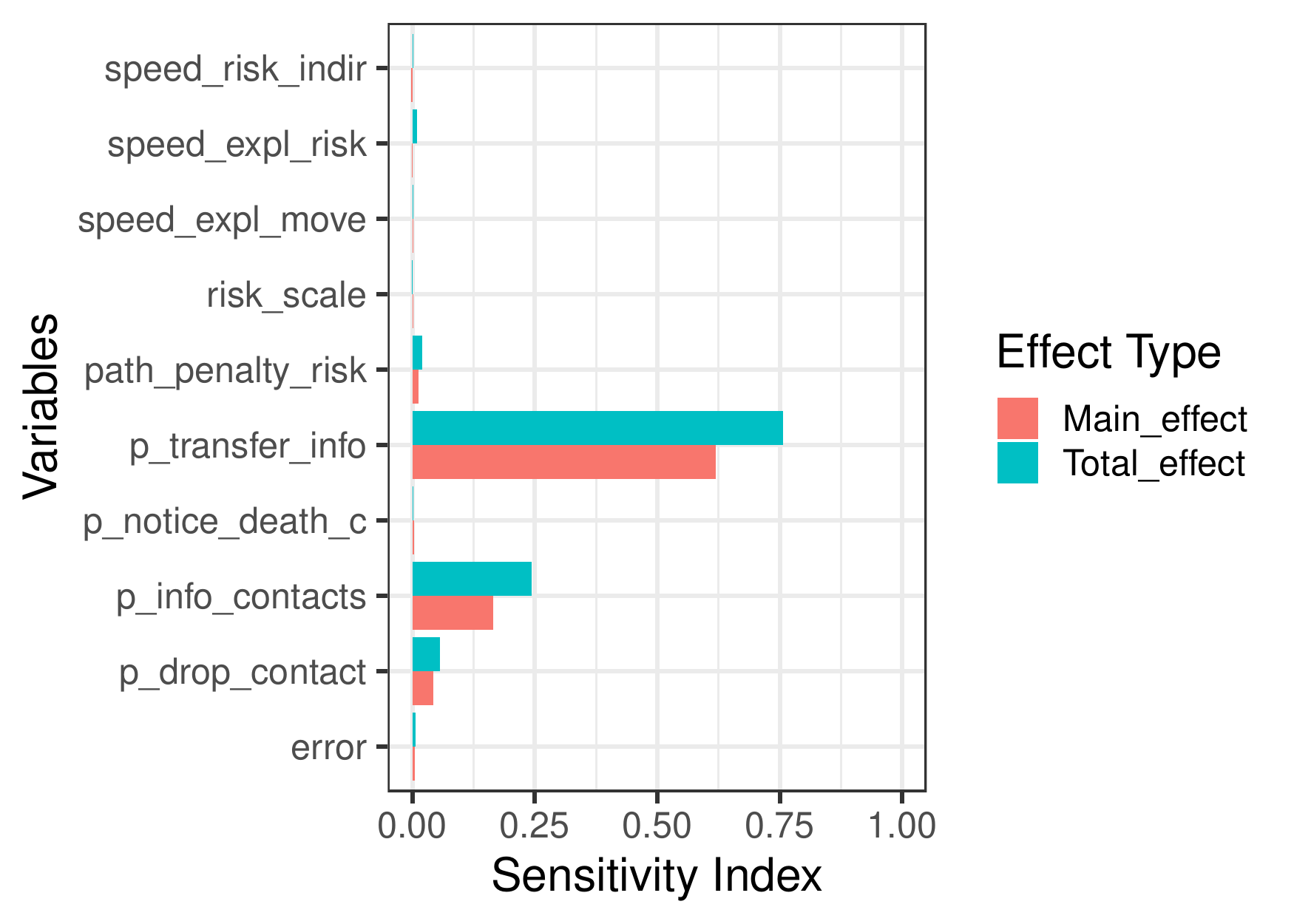}
		\caption{Main and total effects for M4}
	\end{subfigure}
	\caption{Sensitivity indices calculated for the output mean\_freq\_plan with (a) the original experiment and model M3, and (b) the automatically reused experiment and model M4.}
	\label{fig:resultsSensitivity}
\end{figure}

Executing the rule generates a new activity a3, and used-dependencies are drawn to the new simulation model M4, the previous experiment E2, and other input entities involved in a2, i.e., K01 and S1. 
As output, according to the ``sensitivity analysis'' pattern, an experiment entity E3, and a simulation data entity S3 are generated.
When generating the new experiment specification for entity E3, the model name has to be adapted from ``M3'' to ``M4''.
In addition, the list of parameters is updated to include also the factors concerning risk behavior. 
Running the experiment produces main and total order sensitivity information which is added to the information model of S3. 
Figure~\ref{fig:resultsSensitivity} shows the sensitivity data for M3 (left) and M4 (right) for the output mean\_freq\_plan, which captures the proportion of time for which agents are following their plan for transiting the space. 
In both model versions, the parameter p\_transfer\_info was identified as the key driver for planning behavior, indicating that information transfer was crucial for plan choice. 
However, it is noticeable that the addition of risk behavior in M4 does not have any substantial influence on this particular output (although this is not the case for other outputs).

Following the experiment results, the simulation model is refined further.
New data entities and research questions are included in the next model refinement step, which produces the model version M5.
Again, triggered by the refinement activity, our approach automatically repeats the previous sensitivity analysis experiment. 
This time the activity m5 is taken as the anchor point for matching the rule r1.
The activity a4 is automatically generated including the new entities E4 and S4,
and new used-connections to the inputs K01, S1, and the previous experiment E3.

\subsection{Cross-Validation of two Models of the Wnt/\textbeta-Catenin Signaling Pathway}

The Wnt/\textbeta-Catenin signaling pathway is a central pathway in development and homeostasis of cells~\cite{Nusse2008}.
Degenerated forms of this pathway are involved in a number of cancers and neurological disorders~\cite{Clevers2012}.
The study by Haack et al. (2015) analyses the regulation of Wnt signaling in the initial cell fate commitment phase of neuronal progenitor cells~\cite{Haack2015spatio}.
In-vitro experiments indicated that raft- and redox-dependent signaling events play a crucial role in the regulation of Wnt signaling during this phase. 
A previous simulation model of the Wnt signaling pathway by Lee et al. (2003)~\cite{Lee2003roles} does not contain the corresponding model entities to accurately represent these regulatory mechanisms.  
Therefore the Lee model was extended by a membrane model component as well as additional intracellular model entities. 
The extended model was calibrated against new in-vitro data and subsequently cross-validated with simulation data from the Lee study to ensure that the basic model behavior was not changed due to the extension and (re-)calibration of the model. 

\begin{figure}
	\centering
	\includegraphics[width=\textwidth]{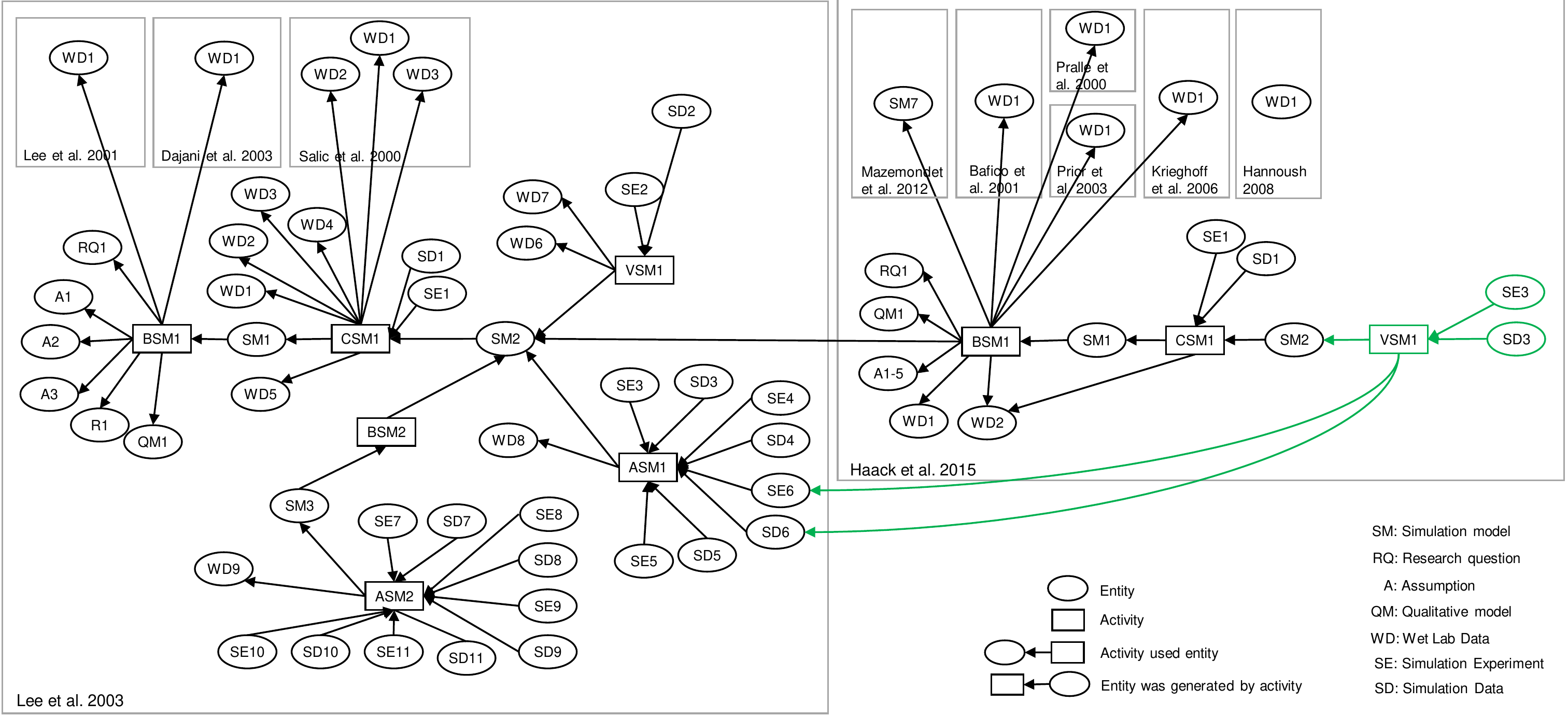}
	\caption{
		Provenance graph showing the simulation study by Lee et al. (2003) and the initial phase of the Haack et al. (2015) study. 
		The both studies are related by the model building activity BSM1. 
		Following the calibration activity CSM1, a cross validation is triggered for the new model SM2. 
		Thus, a new validation activity VSM1 is generated which reuses the simulation experiment SE6 and the simulation data SD6 from the Lee study, and produces an adapted simulation experiment SE3 and the simulation data SD3.}
	\label{fig:haack2015Graph}
\end{figure}

Provenance for both simulation studies was documented by~\cite{Budde2021relating}.
Figure~\ref{fig:haack2015Graph} (left) depicts the entire provenance graph of the Lee study. 
For the Haack et al. study we only show the initial phase with the first few activities (model building BSM1 and calibration CSM1) until the first automatic experiment generation.
The connection between the models is shown by a used relationship between SM2 of the Lee model and BSM1 of the Haack model.
Following the model extension, a calibration experiment is conducted. 
This is recognized as a trigger by our approach\footnote{The resources for the Wnt case study are provided with the main repository: \url{https://git.informatik.uni-rostock.de/mosi/exp-generation}}.

In particular, the cross validation rule $r2$ can be applied at CSM1.
Note that to save space, in Figure~\ref{fig:haack2015Graph} activities of the same type are displayed in aggregated form (e.g., ASM1 of the Lee study aggregates four experiment activities).
Thus, this yields four different matches of the rule.
E.g., $match_1$ matches SE5, SD5, and input data WD8 as suitable candidates for reuse, and $match_2$ matches SE6 and SD6:

\begin{align*}
	match_1 = \{ &SM' \leftarrow SM1, D \leftarrow WD2, X \leftarrow \emptyset, SE' \leftarrow SE1, SD' \leftarrow SD1, SM'' \leftarrow SM2, \\
	&SM \leftarrow SM2, Y \leftarrow \{WD8\}, SE \leftarrow SE5, SD \leftarrow SD5
	\} \\
	match_2 = \{ &SM' \leftarrow SM1, D \leftarrow WD2, X \leftarrow \emptyset, SE' \leftarrow SE1, SD' \leftarrow SD1, SM'' \leftarrow SM2, \\
	&SM \leftarrow SM2, Y \leftarrow \emptyset, SE \leftarrow SE6, SD \leftarrow SD6
	\}.
\end{align*}

Figure~\ref{fig:haack2015Graph} exemplarily shows the generated validation activity based on SE6 and SD6 of the Lee et al. study and the new model SM2 of the Haack et al. study.
The simulation experiment SE6 was specified in SED-ML, and the corresponding model SM2 of the Lee et al. study in SBML~\cite{Hucka2003systems}.
The files\footnote{\url{https://www.ebi.ac.uk/biomodels/BIOMD0000000658}} are available on the BioModels database~\cite{BioModels2020}.
However, as the Wnt model by Haack et al. was specified using the rule-based modeling language ML-Rules~\cite{Helms2017semantics}, and the experiments are conducted using the experiment specification language SESSL~\cite{Ewald2014sessl}, during the adaption and generation step, the experiment specification has to be translated. 

During the translation, to receive an executable experiment specification, it is checked whether the observed species are still known under the same name in the extended model.
Here, we facilitate the qualitative models (both denoted QM1) based on which the simulation models were built.
The qualitative models contain a list of species, each annotated with proteome identifiers using the Uniprot proteome identifier (UPID)~\cite{Uniprot2019}.
Table~\ref{tab:wntOntology} shows a mapping of selected model species from the Lee and Haack models.
In the case of beta-catenin, which is the variable of interest in this experiment, no translation is required.
\begin{table}
	\caption{UniProtKB IDs (prefixed by UniProtKB) are used to identify the species involved in the Lee model, and to map them to species in the Haack model.}
	\label{tab:wntOntology}
	\begin{tabular}{lll}
		\toprule
		Lee Model & Haack Model & Ontology Tag \\
		\midrule
		W & Wnt & UniProtKB - P31285 (WNT3A) \\
		Axin & Axin & UniProtKB - Q9YGY0 (AXIN1) \\
		\textbeta-catenin & \textbeta-catenin & UniProtKB - P26233 (CTNB1)  \\
		Dsh & Dvl & UniProtKB - P51142 (DVL2) \\
		\bottomrule
	\end{tabular}
\end{table}

But also other aspects of the experiments or models might have to be considered when reusing an experiment from a different simulation study.
For instance, in our case study, both models are subject to different time scales. 
This is because the parameters of both models were fitted against experimental data from Xenopus egg extracts and human neural progenitor cells, two cell-biological systems with different time scales.
Therefore, when comparing simulation results between both models, experiment outputs need to be translated by a certain factor.

\begin{figure}
	\centering
	\includegraphics[width=0.6\textwidth]{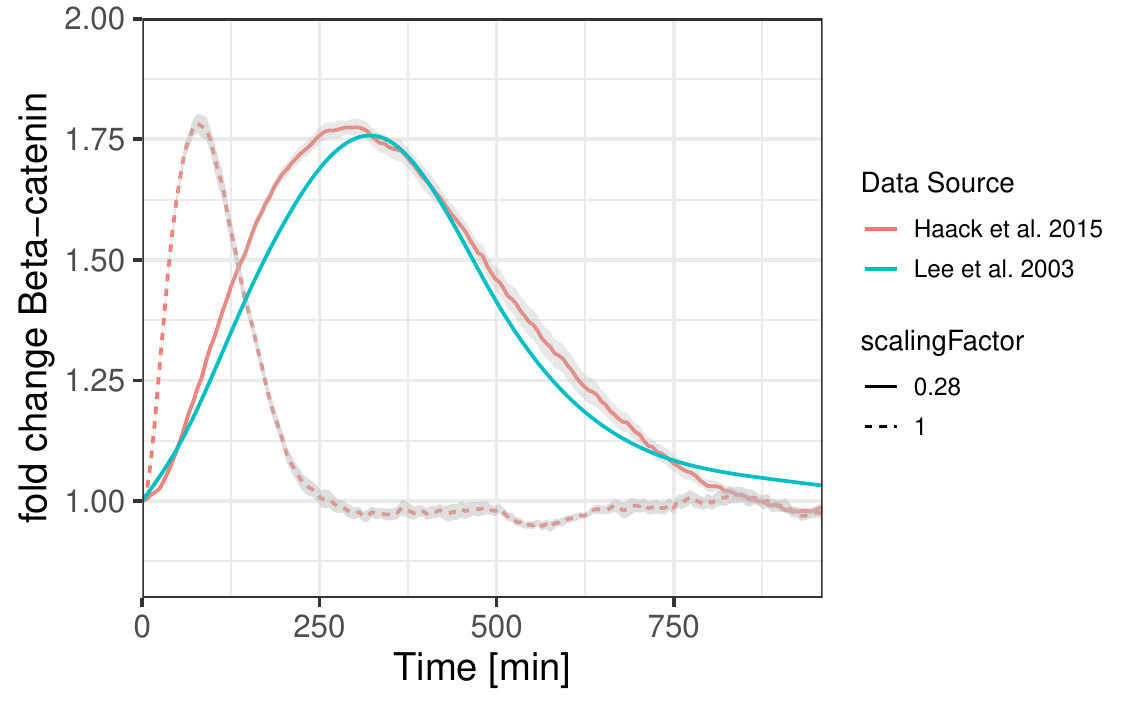}
	\caption{Results of the cross-validation experiment. Beta-catenin values are compared between the Lee model and the Haack model before and after applying the scaling factor.}
	\label{fig:resultsCrossValidation}
\end{figure}

After these adaptations, the new experiment SE3 is generated and can finally be executed.
Figure~\ref{fig:resultsCrossValidation} shows the cross-validation results. 
It compares the trajectories of the key protein \textbeta-catenin, an indicator of the pathway's activity, produced by the Lee and the Haack model (with adapted time scale) when stimulated with a transient stimulus.
After applying the correct transformation factor, that corrects for the varying time scales, the \textbeta-catenin curves show the same maximum at the same time. 
This means that the extensions applied in the study of Haack et al. do not alter the dynamics of the pathway.

\subsection{Discussion}

Changes (or lack of changes) in the sensitivity of simulation outputs to parameters following substantive alterations to the model provide important information about the mechanisms at work in a simulated system. 
In the case of the migration model above, changes in the decision-making process of agents did not hugely affect the output in question. 
The ability to automatically identify the experiment needed for a sensitivity analysis (and subsequently to trigger this experiment following a change in the model) significantly shortens the modeling cycle and reduces the burden on the modeler. 
This means that the focus can be on analyzing and interpreting simulation results and considering additional modeling steps.

In addition to shortening the modeling cycle, automated experiment specification is also valuable to increase the models validity and reusability. 
For instance, in the Haack study, an existing model was extended by further model components that significantly altered the structure of the model. 
However, despite the changes, the model should still be able to reproduce basic dynamics of the pathway that were either obtained in wet-lab experiments or by simulations of a previous model version. 
The ability to automatically identify and reuse all experiments that are necessary to establish validity of a model allows for an easier and more rigorous validation of extended models. 

In both case studies, we successfully demonstrated how provenance information can be exploited in conducting simulation studies in a more systematic and effective manner.
Currently, provenance is only recorded for very few models.
Nonetheless, a variety of meta-information is already documented in domain-specific formats.
Most notably, the systems biology community made substantial progress in standardizing their specification formats as well as the way they package and share simulation models~\cite{Schreiber2020specifications,Waltemath2020combine, Tiwari2021reproducibility}. 
Similarly, in agent-based modeling and empirical studies reporting guidelines are widely used~\cite{Monks2019strengthening, Grimm2020odd}.
These are crucial steps to enable an automatic reuse, adaptation, and execution of simulation experiments for a broader range of simulation studies.

\section{Related Work}
\label{sec:relatedwork}

Following the reproducibility crisis~\cite{Pawlikowski2002credibility}, changes in how simulation studies are conducted (and thus how simulation experiments are carried out) were required.
Therefore, in recent years, simulation researchers investigated the reuse and/or generation of simulation experiments or parts thereof.
Table~\ref{tab:rw} summarizes the related work on reusing and/or generating simulation experiments with regards to their overall objective, the central methodology applied, the experiment types supported, the simulation tools supported, and the scope of the automation (i.e., how the generation process is initiated, and whether the approach can be used within or across simulation studies).
These are compared to the features of the approach presented in this paper (RASE).

\begin{table}
	\scriptsize
	\caption{Comparison of related work on generating simulation experiments. n.a. = not applicable.}
	\label{tab:rw}
	\begin{tabularx}{\linewidth}{Xp{.17\linewidth}p{.12\linewidth}p{.1\linewidth}p{.07\linewidth}p{.1\linewidth}p{.07\linewidth}}
		\toprule
		Publication & Objective & Central Methodology & Experiment Types & Supported Tools & Generation Trigger & Across Studies \\
		\midrule
		Birta and \"{O}zmizrak (\citeyear{Birta1996knowledge}) & Efficient and effective experiment design with minimum number of simulations  & Knowledge base about system behavior & Requirement checking & n.a. & n.a. & no \\
		Teran-Somohano et al. (\citeyear{Teran2015model})  & Generation of factorial designs, e.g., for parameter sweeps & Model-driven engineering & Factorial designs & Repast & GUI-based & no \\
		Yilmaz et al. (\citeyear{Yilmaz2016goal}) & Generation of experiment designs from hypotheses  & Model-driven engineering & Hypothesis testing & n.a. & n.a. & no \\
		Lorig (\citeyear{Lorig2019hypothesis})  & Generation of experiment designs from hypotheses  & Formal hypothesis specification & Hypothesis testing & NetLogo & n.a. & no \\
		Cooper et al. (\citeyear{Cooper2016cardiac}) & Comparison of electrophysiological cell models  & Online database of models, experiments and data & Time courses, steady state analysis & CellML, SED-ML & Submission of a new model & yes \\
		Peng et al. (\citeyear{Peng2016reusing,Peng2017reusing}) & Reuse and adaption of simulation experiments for model extension and composition  & Composition of experiment specifications & Statistical Model Checking & SESSL, MITL & User inputs provided in DSL & yes \\
		Ruscheinski et al. (\citeyear{Ruscheinski2019artifact})  & Guidance for re-running experiments based on artifact milestones & Artifact-based workflow & Statistical model checking  & Flexible & Workflow milestones & no \\
		\midrule
		RASE  & Reuse and adaptation of simulation experiments based on provenance information of the simulation study  & Provenance patterns, Model-driven engineering & Flexible & Flexible & Various model building activities & yes \\
		\bottomrule
	\end{tabularx}
\end{table}

Early work by Birta and \"{O}zmizrak already discussed the use of prestructured knowledge for the validation of simulation models~\cite{Birta1996knowledge}.
Their framework focused on efficient experiment designs for checking behavioral requirements. 
The generation of experiment designs was also the topic of the work by Teran-Somohano et al.~\cite{Teran2015model}.
There, model-driven engineering (MDE) was applied for creating various factorial designs through a specialized graphical user interface.
Yilmaz et al. demonstrated the potential of MDE for generating experiment designs for hypothesis testing as part of a goal-hypothesis-experiment framework~\cite{Yilmaz2016goal}.
For generating experiment designs from hypotheses, the formal specification of hypotheses using a domain-specific language (DSL) is central, as shown by Lorig~\cite{Lorig2019hypothesis}.

The cardiac electrophysiology wet lab designed by Cooper et al.~\cite{Cooper2016cardiac} targeted the comparison of simulation models.
An online database stores SED-ML simulation experiment specifications, e.g., for time course or steady state analyses.
When a new model is uploaded to the web application, it is automatically cross-checked with other models of cardiac electrophysiology by running all available experiments on the new model.
Rerunning experiments can also be supported within artifact-based workflows~\cite{Ruscheinski2019artifact}.
Based on a declarative life cycle model, the workflow can guide users to repeating simulation experiments for validation or calibration if certain milestones are achieved or invalidated.  
The concept of toolboxes enables the use of different modeling and simulation tools inside the workflow.
Reusing and adapting simulation experiments is crucial when simulation models are extended or composed.
Peng et al.~\cite{Peng2016reusing, Peng2017reusing} reuse statistical model checking experiments and combine them according to the model composition. 
Logic formulas, specified in Metric Interval Temporal Logic (MITL)~\cite{Maler2004monitoring}, are merged to express and check properties of the composed model.

Our approach adds to the existing research, by providing a provenance-based mechanism that automatically reuses and adapts simulation experiments during the simulation study.
Provenance patterns encapsulate knowledge about the various model building and experimentation activities, and serve as triggers to the experiment generation, and for automatically selecting suitable simulation experiments to reuse.
In addition, provenance contains valuable information about a simulation study that allows for an automatic adaption of the reused experiments.
Integrating this with an MDE approach for generating and translating simulation experiments allows us to flexibly support a) a variety of experiment types and b) a variety of modeling and simulation tools. 
The latter also facilitates the automatic reuse across simulation studies.

\section{Conclusions}
\label{sec:conclusions}
The repetition of simulation experiments forms a salient feature of the model development process.
Regression testing of successive model refinements and cross-validation of related models are just some examples of this.
In this paper, we exploit provenance information to automatically identify suitable experiments to reuse depending on the last model development steps, and to automatically adapt the experiment specifications accordingly for the new model version. 
The central methods of our approach are the definition of provenance patterns and the construction of reuse rules.
Our collection of patterns and rules captures knowledge about simulation studies explicitly, and thereby contributes to the conduction of more systematic and effective simulation studies.
We successfully demonstrated the applicability of our approach in the context of human migration models as well as cell biology models.
The software prototype of our Reuse and Adapt framework for Simulation Experiments (RASE) is publicly and permanently available.

Future use cases of our framework could include multi-user simulation studies.
For instance, when one user completes a validation, this could trigger a new analysis experiment for another submodel, developed by a different user.
Moreover, our patterns could be applied for consistency management during provenance recording. 
When entering new application areas, also the usability and customizability of our framework could be a concern of future work, e.g., by developing a GUI for specifying the patterns.
So far, our framework concentrates on the reuse of existing simulation experiments.
However, another interesting step forward will be exploiting provenance for generating new experiments from scratch (i.e., without reusing a previous experiment specification as blueprint).

\section*{Acknowledgements}
	\textbf{Funding:} This research was funded by the German Research Foundation (DFG) via the grants 320435134 (P.W.), 258560741 (P.W.) and SFB 1270/1–299150580 (F.H.), and by the European Research Council (ERC) via grant CoG-2016-725232 (J.H.). \newline
	\textbf{Author contributions\footnote{Contributor Roles Taxonomy (CRediT) \url{http://credit.niso.org/}}:} Conceptualization: P.W., A.M.U.; Methodology: P.W., A.W., A.M.U.; Software: A.W., P.W., J.H., F.H.; Investigation: P.W., J.H., F.H.; Data Curation: J.H., F.H.; Writing original draft: P.W., A.M.U., Writing review and editing: P.W., A.M.U., J.H., F.H., A.W.; Visualization: P.W., J.H., F.H.; Supervision: A.M.U.; Project administration: P.W.; Funding Acquisition: A.M.U. \newline
	The authors would like to thank Marcus Dombrowsky for his early work on the software prototype.

\bibliographystyle{unsrtnat}
\bibliography{patterns}

\end{document}